\documentclass[aps,prl,reprint,floatfix]{revtex4-2}

\usepackage{amsmath}
\usepackage[separate-uncertainty]{siunitx}
\usepackage{graphicx}
\usepackage{xcolor}

\graphicspath{{figures/}}

\DeclareSIUnit{\Molar}{M}
\DeclareSIUnit{\mM}{\milli\Molar}

\newcommand{\avg}[1]{\left< #1 \right>}
\newcommand{\abs}[1]{\left\vert #1 \right\vert}
\renewcommand{\vec}[1]{\boldsymbol{#1}}

\begin{document}

\title{Worm-like emulsion droplets}

\author{Jatin Abacousnac}
\author{Wenjun Chen}
\author{Jasna Brujic}
\author{David G. Grier}

\affiliation{Department of Physics and Center for Soft Matter
  Research, New York University, New York, NY 10003, USA}

\begin{abstract}
  Forming an interface between immiscible fluids incurs a free-energy cost that usually
  favors minimizing the interfacial area.
  An emulsion droplet of fixed volume therefore
  tends to form a sphere, and pairs of droplets
  tend to coalesce.
  Surfactant molecules adsorbed to the droplets' surfaces
  stabilize emulsions by providing a kinetic barrier to coalescence.
  Here, we show that the bound surfactants' osmotic pressure also competes with the
  droplet's intrinsic surface tension and can reverse the sign
  of the overall surface free energy.
  The onset of negative surface tension favors
  maximizing surface area and therefore favors elongation
  into a worm-like morphology.
  Analyzing this system in the Gibbs grand canonical ensemble reveals
  a phase transition between spherical and worm-like emulsions
  that is governed by the chemical potential of surfactant
  molecules in solution.
  Predictions based on this model agree with the
  observed behavior of an experimental model system composed of
  lipid-stabilized silicone oil droplets in an aqueous
  surfactant solution.
\end{abstract}

\maketitle

\begin{figure}
  \centering
  \includegraphics[width=0.75\columnwidth]{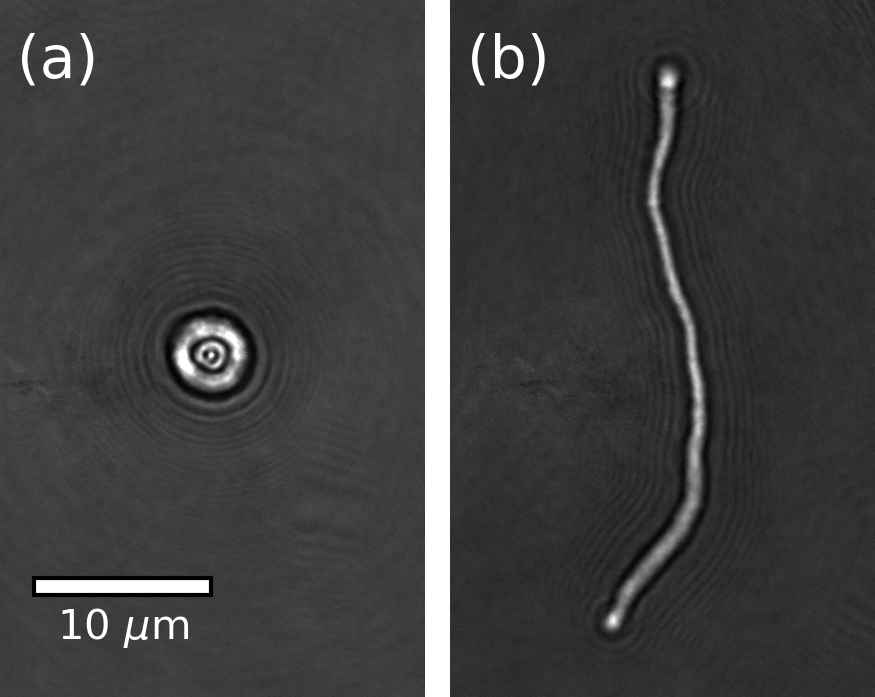}
  \caption{Holographic images of a lipid-stabilized emulsion droplet
  composed of a ternary oil mixture, dispersed in
  an aqueous SDS solution at \qty{1}{\mM}:
  (a) in the spherical morphology
  at \qty{23(1)}{\celsius} and
  (b) in the flexible worm-like morphology
  at a slightly lower temperature, \qty{21(1)}{\celsius}.}
  \label{fig:droplets}
\end{figure}

The droplets in oil-in-water emulsions generally
minimize their interfacial area to lower their free energy. This mechanism favors the formation of spherical
droplets and promotes droplet coalescence.
Droplets can be kinetically stabilized against
coalescence by binding surfactant molecules to the
oil-water interface.
Here, we describe an additional dramatic consequence
of surfactant binding, illustrated in Fig.~\ref{fig:droplets}, in which
the osmotic pressure of bound molecules
drives a spherical emulsion droplet through a
phase transition to a worm-like morphology.
Worm-like emulsions are distinct from worm-like micelles
\cite{cates1990statics,dreiss2007wormlike}
because they are stabilized by surfactant concentrations
far below the critical micelle concentration (CMC).
They also differ from the varied morphologies
of liposomes and vesicles, which are dictated by
the bending and curvature moduli of surfactant
bilayers
\cite{zidovska2021membrane}.
The sphere-to-worm transition in suitable emulsions
constitutes an alternative pathway
for morphology control in soft matter.

Our study of worm-like emulsions is inspired by
the observed behavior of micrometer-scale
droplets of a ternary oil mixture stabilized by
lipid surfactants and dispersed in water \cite{chen2024refractive}.
This model system has been studied extensively as
a building block of emulsion-droplet chains
known as colloidomers
\cite{mcmullen2018freely}.
Individual droplets ordinarily form spheres, as shown
in Fig.~\ref{fig:droplets}(a).
We find that adding a cosurfactant,
such as sodium dodecyl sulfate (SDS),
can cause these droplets to elongate into flexible
worm-like structures
such as the example in Fig.~\ref{fig:droplets}(b).
Spherical droplets are observed to
deform continuously into worms
when their dispersion is cooled a few degrees,
and reform into spheres upon heating.
The transformation is reversible through
repeated cycles.
Simple linear worms result from
quasistationary cooling and more complex morphologies
emerge at higher cooling rates.

\begin{figure}
  \centering
  \includegraphics[width=0.75\columnwidth]{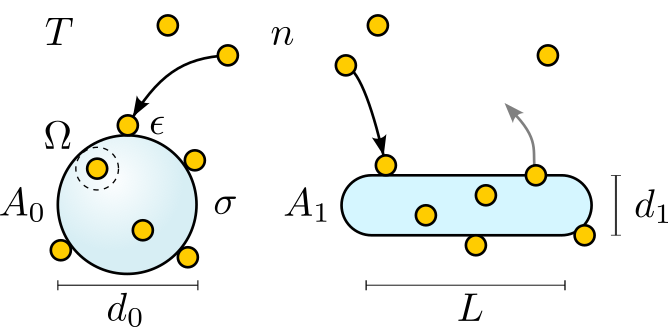}
  \caption{Schematic representation of surfactant molecules
    (yellow) binding to a spherical emulsion droplet, thereby
    inducing it to adopt a worm-like morphology.}
  \label{fig:schematic}
\end{figure}

The observed phenomenology of worm-like emulsions
is captured
by the minimal model depicted schematically in Fig.~\ref{fig:schematic}.
This model treats an oil droplet as the substrate for
a two-dimensional classical gas of surfactant molecules.
Each surfactant molecule has mass $m$ and subtends an area $\Omega$ on the droplet's surface.
Surfactant molecules bind to the surface
with a binding energy, $\epsilon$, that we assume to be independent of coverage.
A system of $N$ bound molecules
therefore has the Hamiltonian
\begin{equation}
  \label{eq:hamiltonian}
  H_N = \sum_{n = 1}^N \frac{p_n^2}{2m} - N \epsilon,
\end{equation}
where $\vec{p}_n$ is the two-dimensional momentum of the
$n$-th molecule on the droplet's surface.
The associated canonical partition function is
\begin{subequations}
  \label{eq:canonicalpartitionfunction}
  \begin{align}
    Z(A, T, N)
    & =
      \frac{1}{N!} \frac{1}{h^{2N}} \int
      e^{- \beta H_N} \,
      \prod_{j = 1}^N d^2 r_j \, d^2 p_j
       \\
    & =
      \frac{1}{N!} \frac{1}{A_Q^N} \left(A - \frac{1}{2} N\Omega
      \right)^N \, e^{N \beta \epsilon},
      \label{eq:excludedarea}
  \end{align}
\end{subequations}
where $h$ is Planck's constant,
$\beta^{-1} = k_B T$ is the thermal energy scale
at absolute temperature $T$,
$A$ is the droplet's surface area, and
\begin{subequations}
  \label{eq:quantumarea}
  \begin{align}
    A_Q(T)
    & =
      \left[\frac{4 \pi}{h^2} \int_0^\infty p \, \exp\left(- \beta \frac{p^2}{2m} \right) \, dp \right]^{-1}\\
    & =
      \frac{h^2}{2 \pi m \, k_B T}
  \end{align}
\end{subequations}
is the quantum area for the bound molecules.
The
term in parentheses on the right-hand side of Eq.~\eqref{eq:excludedarea} is
the effective area of the droplet's surface after accounting for the
area occupied by surfactant molecules.

Bound surfactant molecules are in thermodynamic equilibrium with
the bulk solution of free surfactant molecules at chemical potential $\mu$.
In establishing this equilibrium, the osmotic pressure of the
molecules on the surface favors increasing
the droplet's surface area by deforming the droplet into
a noncompact shape such as a worm.
Increasing the surface area, however, incurs a cost in surface energy
through the fluid's intrinsic surface tension, $\sigma$.
We account for changes in  the droplet's
surface area and surfactant
coverage with the Gibbs grand canonical partition function,
\begin{align}
    \label{eq:ggcpf}
  \mathcal{Q}(\sigma, \mu, T)
  & =
    \int_{A_0}^{A_1} \frac{dA}{A_Q} \, \sum_{N = 0}^{M(A)}
    Z(A, T, N) \,
    e^{\beta N \mu} e^{-\beta \sigma A},
\end{align}
where $A_0 = \pi d_0^2$ is the surface area of the bare spherical droplet at
diameter $d_0$, and $A_1$ is the maximum area of the deformed worm-like droplet.
The minimum area is set by the
droplet's volume.
The maximum area is limited by factors
such as the bending energy of the lipid
surfactant \cite{cates1990statics}
whose microscopic mechanisms we do not
address in the present study.
Assuming that surfactant molecules do not overlap,
close packing limits the coverage to
$N \leq M(A) = A/\Omega$.

Equation~\eqref{eq:ggcpf} can be recast as
\begin{equation}
  \mathcal{Q}(\sigma, T, \mu)
  =
    \int_{A_0}^{A_1} \frac{dA}{A_Q} \,  \mathcal{Z}(A, T, \mu)  \, e^{-\beta \sigma A} ,
\end{equation}
in terms
of the grand canonical partition function,
\begin{equation}
\label{eq:Z}
  \mathcal{Z}(A, T, \mu)
  =
    \sum_{N = 0}^{M(A)}
    \frac{1}{N!} \, \left(\frac{A - \frac{1}{2} N \Omega}{A_Q}\right)^N e^{\beta N (\epsilon + \mu)} .
\end{equation}
For simplicity, we assume that the coverage is small enough,
$\avg{N} \ll A/\Omega$,
that we may neglect the excluded area.
In this approximation, the grand canonical partition function
reduces to
\begin{subequations}
  \label{eq:grandpartitionfunction}
  \begin{align}
    \mathcal{Z}(A, T, \mu)
    & \approx
      \sum_{N = 0}^\infty
      \frac{1}{N!} \left[ \frac{A}{A_Q} \, e^{\beta(\epsilon + \mu)} \right]^N \\
    & =
      \exp\left( \frac{A}{A_Q} e^{\beta(\epsilon + \mu)} \right).
  \end{align}
\end{subequations}
The Gibbs grand canonical partition function then follows as
\begin{subequations}
  \label{eq:gibbsgrandcanonicalpartitionfunction}
  \begin{align}
    \mathcal{Q}(\sigma, T, \mu)
    & \approx
      \int_{x_0}^{x_1}
      \exp\left( x \left[
      e^{\beta(\epsilon + \mu)} - \beta \sigma A_Q \right] \right) \, dx \\
    & =
      \frac{e^{-\kappa x_0} - e^{-\kappa x_1}}{\kappa},
  \end{align}
\end{subequations}
where $x = A/A_Q$ is the dimensionless surface area and
\begin{equation}
  \label{eq:surfacetension}
  \kappa(T)
  =
  \beta \sigma A_Q(T) - e^{\beta(\epsilon + \mu)}
\end{equation}
is a dimensionless parameter that describes the effective surface tension.
The sign of $\kappa(T)$ selects the droplet
morphology and
is determined by
the temperature and by the
chemical potential of the surfactant
molecules.

\begin{figure}[!t]
  \centering \includegraphics[width=0.9\columnwidth]{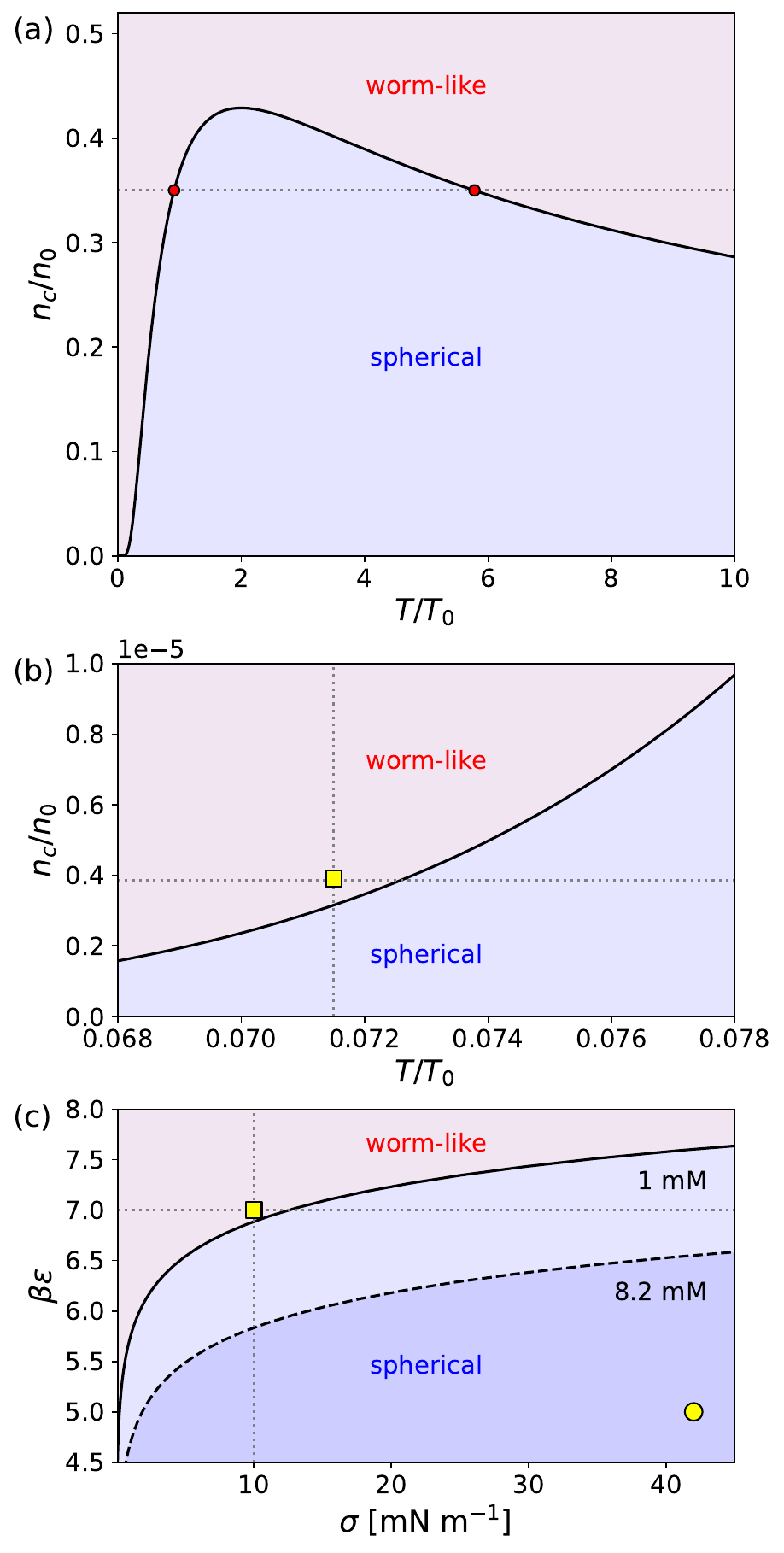}
  \caption{(a) Phase boundary between spherical
  and worm-like morphologies predicted by Eq.~\eqref{eq:phaseboundary} as a function
    of temperature, $T$, and surfactant concentration, $n$.
  (b) Region of the phase diagram around the experimental
  conditions in Fig.~\ref{fig:droplets}, which are
  plotted as a discrete (square) point.
  Proximity to the predicted phase boundary is
  consistent with the system's sensitivity
  to temperature changes.
  (c) Phase boundary plotted in terms of the
  intrinsic interfacial energy, $\sigma$, of the droplets
  and the binding energy for SDS, $\epsilon$, in an
  SDS solution at \qty{1}{\mM} (solid curve) and at
  the CMC for SDS (dashed curve). Properties for the ternary oil
  droplets used for this study fall near
  the phase boundary at \qty{1}{\mM} (square).
  Properties for standard silicone oil in SDS solution \cite{kanellopoulos1971adsorption}
  (circle) lie far from the phase boundary,
  consistent with observation.}
  \label{fig:phasediagram}
\end{figure}

Treating unbound sufactant molecules as an ideal gas with a bulk number density $n$, the
chemical potential is given by the Sackur-Tetrode formula \cite{sackur1911anwendung, tetrode1912chemische},
\begin{equation}
  \label{eq:chemicalpotential}
  \mu
  = k_B T \, \ln\left( n V_Q \right) ,
\end{equation}
where $V_Q = A_Q^{\frac{3}{2}}$ is the quantum volume
for the molecules in the bulk.
Applying this to Eq.~\eqref{eq:surfacetension}
suggests that the effective surface tension vanishes along a
curve,
\begin{equation}
  \label{eq:phaseboundary}
    \frac{n_c(T)}{n_0}
    =
    \left(\frac{T_0}{T}\right)^{\frac{1}{2}} \,
    \exp\left(-\frac{T_0}{T} \right) ,
\end{equation}
that serves as the phase boundary between
the spherical and worm-like morphologies.
This curve is plotted in Fig.~\ref{fig:phasediagram}(a).
The temperature scale in Eq.~\eqref{eq:phaseboundary}
is set by the surfactant's binding
energy through
\begin{equation}
    T_0 = \frac{2 \epsilon}{k_B} ,
\end{equation}
while the characteristic concentration scale,
\begin{equation}
    \quad n_0 =
    \sqrt{\frac{2 \pi m}{h^2} \frac{\sigma^2}{\epsilon}} ,
\end{equation}
balances binding energy against surface tension.

The effective surface tension is positive in the absence of dissolved
surfactant molecules,
$n = 0$, which is consistent with the expectation that
emulsion droplets ordinarily form spheres.
Conversely, $\kappa(T)$ is strictly negative when
the concentration of dissolved surfactant molecules exceeds
$n^\ast = n_0/\sqrt{2e} = \num{0.429} \, n_0$.
Between these two
limiting cases, the sign of the effective surface tension depends on
the temperature,
with spherical droplets being stable in a range,
$T_- < T < T_+$, that is set by the
bulk surfactant concentration.

The average surface area adopted by a droplet can be
obtained in the Gibbs grand canonical ensemble,
\begin{subequations}
  \begin{align}
    \avg{A}
    & =
      - \frac{1}{\beta} \frac{\partial}{\partial \sigma} \ln \mathcal{Q} \\
    & =
      \frac{A_Q}{\kappa}
      \frac{(\kappa x_0 + 1) \, e^{-\kappa x_0} - (\kappa x_1 + 1) \, e^{-\kappa x_1}}{
      e^{-\kappa x_0} - e^{-\kappa x_1}}.
  \end{align}
\end{subequations}
Assuming that $\abs{\kappa} x_1 \gg \abs{\kappa} x_0 \gg 1$, the sign
of $\kappa$ selects either spheres or worms:
\begin{equation}
  \avg{A}
  \approx
  \begin{cases}
    A_0, & T_- < T < T_+ \Rightarrow \text{spherical}\\
    A_1, & \text{otherwise} \Rightarrow \text{worm-like}
  \end{cases} .
\end{equation}

The nature of the surfactant-mediated
phase transition is clarified by
considering the number of surfactant
molecules that are bound to the
droplet:
\begin{equation}
  \label{eq:occupation}
  \avg{N}
  =
  \frac{1}{\beta} \frac{\partial \ln \mathcal{Q}}{\partial \mu}
  =
  \frac{\avg{A}}{A_Q} \, e^{\beta(\epsilon + \mu)} .
\end{equation}
This corresponds to a fractional surface coverage,
\begin{align}
\label{eq:coverage}
  \phi
  \equiv \Omega \frac{\avg{N}}{\avg{A}}
  =
    n \, \Omega \, A_Q^{\frac{1}{2}} \, e^{\beta \epsilon} ,
\end{align}
that depends strongly on temperature.
High surfactant coverage in the low-temperature phase
provides the osmotic pressure needed to overcome
the droplet's intrinsic surface tension and
destabilize the spherical morphology.
Equation~\eqref{eq:coverage} also can be used
to assess whether or not
the surfactant's excluded area can be neglected
in deriving Eq.~\eqref{eq:grandpartitionfunction}.
Taking the excluded area into account
tends to reduce the number of bound surfactant
molecules while
also tending to increase their surface pressure.
These competing effects
influence the phase boundary, $n_c(T)$,
quantitatively but not qualitatively.

The predicted sphere-to-worm
transition can be compared
directly to the observed
behavior of emulsion droplets.
The droplets used in this study
are synthesized from a 1:1:1 mixture by volume of three silane monomers: (a) dimethyldiethoxysilane (Sigma-Aldrich), (b) (3,3,3-
trifluoropropyl) methyldimethoxysilane (Gelest), and
(c) 3-glycidoxypropylmethyldiethoxysilane (Gelest).
They are formed into
micrometer-scale droplets by ammonia-catalyzed hydrolysis
followed by condensation, as described in \cite{elbers2015bulk, mcmullen2022self}.
The droplets are cleaned and dispersed in \qty{1}{mM} SDS solution
to an overall concentration of \qty{1e7}{droplets\per\milli\liter} before being
stabilized by adding
DSPE-PEG-SH lipid (Avanti Polar Lipids, MW 2000) at
a concentration of \qty{5}{\micro\gram\per\milli\liter}.
This procedure yields droplets with
an estimated surface coverage of lipids of
roughly \qty{50}{percent} \cite{mcmullen2022self}.
Lipids bind irreversibly to the oil-water
interface and are not
observed to form micelles.

The \qty{1}{\mM} concentration of SDS in our model system
phenomenologically is close to the
point at which spherical droplets deform into worms.
The transformation is not observed
in dispersions that lack SDS.
Each SDS molecule has a mass
of $m = \qty{4.8e-25}{\kilo\gram}$ and subtends
$\Omega \lesssim \qty{0.5}{\square\nm}$
on the oil-water interface
\cite{deaguiar2010interfacial}.
The interfacial tension is
measured with a pendant drop tensiometer
(attension, Theta Lite) to be
$\sigma = \qty{10}{\milli\newton\per\meter}$
in the absence of SDS,
which is consistent with values obtained
for similar systems in previous studies
\cite{brujic20033d,lin2016evidence}.
The dependence of the surface energy on SDS concentration
yields $\epsilon = \qty{7}{k_B T}$
for the surfactant binding energy
\cite{lin2016evidence}.
Equation~\eqref{eq:coverage} then
yields $\phi \lesssim \num{2e-3}$
for the
fractional surface coverage of SDS at room temperature,
which is small enough
to justify neglecting $\Omega$ in the derivation of
$\mathcal{Z}(A, T, \mu)$, but still large
enough to drive the transition.
Figure~\ref{fig:phasediagram}(b) confirms that
the model system's properties
lie close to the predicted phase boundary.

Figure~\ref{fig:phasediagram}(c) recasts these
results in terms of the surface tension
and binding energy at the transition point
given the molar mass and concentration of the
surfactant.
This again confirms that the material properties
of the model system are close to the predicted
transition point in a \qty{1}{\mM} SDS solution
at room temperature.
Conventional silicone oils, by contrast, have
substantially higher surface tension in contact with water and
lower affinity for SDS \cite{kanellopoulos1971adsorption},
and therefore lie well within the predicted
domain of stability for spherical droplets.
Such systems cannot be destabilized into worm-like
droplets by increasing the surfactant
concentration because the CMC for SDS
at \qty{8.2}{\mM}
establishes an upper bound on the chemical
potential of bound molecules at the interface.
This may explain why worm-like emulsions
have not been reported previously.

We analyze emulsion droplets' behavior using a custom-built instrument
that combines conventional bright-field microscopy,
in-line holographic microscopy
and holographic optical trapping
\cite{obrien2019above}.
The microscope uses an oil-immersion objective lens
(Nikon S Plan Apo, 100\texttimes\ numerical aperture 1.4)
to achieve a system magnification of \qty{48}{\nm\per pixel} on the
sensor of a CMOS video camera (Flir Flea3 USB3.0).
Holographic imaging is performed
at a vacuum wavelength of \qty{447}{\nm}
(Coherent Cube) and
an intensity of \qty{1}{\milli\watt\per\square\mm}.
Holographic trapping is performed at
\qty{1070}{\nm} (IPG Photonics YLR-10-LP)
using a phase-only spatial light modulator
(Holoeye Pluto).
The images in Fig.~\ref{fig:droplets} are holograms of the emulsion droplet
that can be used to numerically reconstruct the deformed
droplet's three-dimensional
structure \cite{dixon2011holographic} and to track its
motion \cite{abacousnac2024measuring}.

\begin{figure}
    \centering
    \includegraphics[width=\columnwidth]{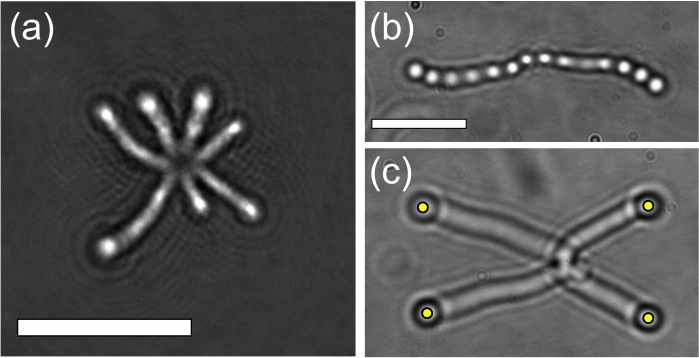}
    \caption{(a) Holographic image of a multibranched worm-like emulsion formed by
    rapid cooling
    (b) Bright-field image of a worm-like
    emulsion droplet undergoing a pearling instability.
    (c) Bright-field image of two worm-like emulsion droplets crossed and placed under tension by optical tweezers
    (yellow circles) projected onto their ends.
    Scale bars: \qty{10}{\um}.}
    \label{fig:variants}
\end{figure}

The ramified structure in Fig.~\ref{fig:variants}(a) is a worm-like
emulsion droplet formed from a spherical droplet by cooling
the microscope stage
\qty{3}{\celsius}
in \qty{1}{\minute}. Rather than continuously elongating into a uniaxial worm,
this rapidly cooled droplet nucleated seven worm-like extensions.
Even though the individual arms are flexible
they do not relax into a uniaxial worm
resembling Fig~\ref{fig:droplets}(b) presumably because surface
osmotic pressure tends to keep them all extended.
The formation and stability of such star-like worms therefore
provides additional support for the proposed morphology selection mechanism.
Multi-arm droplets contract into spheres upon heating
and can be thermally cycled into either uniaxial or ramified worms
through the choice of cooling rate.

Rapidly heating a linear worm can induce a pearling instability of the kind
depicted in the bright-field image in Fig.~\ref{fig:variants}(b).
In this case, the sample's temperature was raised by \qty{3}{\celsius}
over \qty{1}{minute}, which places it roughly on the predicted phase
boundary.
The pearling instability does not cause the droplet to break up
but does appear to hinder its return to sphericity.
The delay may be caused by stress-induced redistribution of lipid surfactants on the droplet's
surface, which would require time to relax.

The image in Fig.~\ref{fig:variants}(c) highlights worm-like droplets' stability against coalescence.
Two droplets are trapped by holographic optical tweezers
that are projected onto their ends.
The droplets' axes are made to intersect by moving the
traps interactively in three dimensions.
The droplets then are placed under
an estimated \qty{100}{\femto\newton} of tension by moving the tweezers
apart.
Each micrometer-diameter
trap is powered by \qty{10}{\milli\watt}, and
the intense illumination raises the local
temperature by roughly \qty{1}{\celsius}.
This localized heating causes the ends of
the worms to form spherical bulges that are
visible in Fig.~\ref{fig:variants}(c).
Despite \qty{2}{\minute} of heating and the imposed
tension, the two worms do not merge.
Extinguishing the optical tweezers allows them
to separate and to diffuse independently.

The observations reported in
Fig.~\ref{fig:variants} lend credence to the
proposal that the sphere-to-worm transformation
is a phase transition and that the
worm-like morphology is
stabilized by the surface osmotic pressure
of bound surfactant molecules.
They further demonstrate that the nature of
the morphological
transformation can be controlled kinetically,
with different morphologies appearing at different cooling and heating
rates.
The discovery of worm-like emulsions in our model system raises
the possibility that similar shape-shifting transformations
will be found in other emulsions.
The ease with which the sphere-to-worm transformation
can be controlled both macroscopically and microscopically
suggests that this class of emulsions will be useful
for developing active and responsive soft materials \cite{tang2015stimuli,liu2020responsive,shen2020stimuli,apsite2021materials}.

We are grateful to Alexander Y.\ Grosberg, Jacques Fattaccioli, Paul Chaikin, and Aditya Hardikar for helpful conversations.
This work was supported by the National Science Foundation through
Awards No.\ DMR-2104837 and DMR-2105255 and the Swiss National Science Foundation through Grant No.\ 10000141.
The integrated instrument for holographic
trapping and microscopy was constructed
under the MRI program of the NSF
through Award No.\ DMR-0922680.


%

\end{document}